\documentstyle[12pt,psfig]{article}
\newcommand{\gtrsim}{\mathop{>}\limits_{\displaystyle{\sim}}}
\newcommand{\tr}{\mbox{tr}}
\def\fsl#1{\setbox0=\hbox{$#1$}           
   \dimen0=\wd0                                 
   \setbox1=\hbox{/} \dimen1=\wd1               
   \ifdim\dimen0>\dimen1                        
      \rlap{\hbox to \dimen0{\hfil/\hfil}}      
      #1                                        
   \else                                        
      \rlap{\hbox to \dimen1{\hfil$#1$\hfil}}   
      /                                         
   \fi}                                         %
\makeatletter

\def\@aabuffer{}
\def\author #1{\expandafter\def\expandafter\@aabuffer\expandafter
{\@aabuffer \small\rm      #1\relax \par}}
\def\address#1{\expandafter\def\expandafter\@aabuffer\expandafter
{\@aabuffer \small\it #1\relax \par\vspace{1em}}}

\def\maketitle{\newpage
\def\thefootnote{\fnsymbol{footnote}}
 \null
 {\normalsize \tt \begin{flushright} 
  \begin{tabular}[t]{l} \@date 
  \end{tabular}
 \end{flushright}}
\begin{center}
   {\bf \@title \par}       
   \vskip 2em                      
   \@aabuffer\relax
\end{center} \par
\gdef\@aabuffer{}
\setcounter{footnote}{0}
\def\thefootnote{\alph{footnote}}
}

\def\abstracts#1{
\begin{center}
{\begin{minipage}{4.2truein}
                 \footnotesize
                 \parindent=0pt #1\par
                 \end{minipage}}\end{center}
                 \vskip 2em \par}

\fussy
\def\section{\@startsection {section}{1}{\z@}{-3.5ex plus -1ex minus 
    -.2ex}{2.3ex plus .2ex}{\bf }}
\def\subsection{\@startsection{subsection}{2}{\z@}{-3.25ex plus -1ex minus 
   -.2ex}{1.5ex plus .2ex}{\it }}

\def\thefootnote{\alph{footnote}}
\def\@makefnmark{{${}^{\@thefnmark}$}}

\renewenvironment{thebibliography}[1]
        {\begin{list}{\arabic{enumi}.}
        {\usecounter{enumi}\setlength{\parsep}{0pt}
         \setlength{\itemsep}{0pt} 
         \settowidth
        {\labelwidth}{#1.}\sloppy}}{\end{list}}

\newcounter{arabiclistc}

\def\@citex[#1]#2{\if@filesw\immediate\write\@auxout
        {\string\citation{#2}}\fi
\def\@citea{}\@cite{\@for\@citeb:=#2\do
        {\@citea\def\@citea{,}\@ifundefined
        {b@\@citeb}{{\bf ?}\@warning
        {Citation `\@citeb' on page \thepage \space undefined}}
        {\csname b@\@citeb\endcsname}}}{#1}}

\newif\if@cghi
\def\cite{\@cghitrue\@ifnextchar [{\@tempswatrue
        \@citex}{\@tempswafalse\@citex[]}}
\def\citelow{\@cghifalse\@ifnextchar [{\@tempswatrue
        \@citex}{\@tempswafalse\@citex[]}}
\def\@cite#1#2{{$^{#1}$\if@tempswa\typeout
        {IJCGA warning: optional citation argument 
        ignored: `#2'} \fi}}

\def\baselinestretch{0.975}
\ifx\selectfont\undefined
\@normalsize\else\let\glb@currsize=\relax\selectfont
\fi

\ifx\selectfont\undefined
\def\@singlespacing{%
\def\baselinestretch{1}\ifx\@currsize\normalsize\@normalsize\else\@currsize\fi%
}
\else
\def\@singlespacing{\def\baselinestretch{1}\let\glb@currsize=\relax\selectfont}
\fi

\long\def\@makecaption#1#2{
   \vskip 10pt 
   \setbox\@tempboxa\hbox{\footnotesize #1: #2}
   \ifdim \wd\@tempboxa >\hsize   
       \footnotesize #1: #2\par   
     \else                        
       \hbox to\hsize{\hfil\box\@tempboxa\hfil}  
   \fi}

\makeatother
\begin{document}
\date{
  KEK TH-462\\
  KEK preprint 95-197\\
  January 1996}
\title{
  $WW$ resonance and chiral Lagrangian\footnote{
    Talk presented by M. Tanabashi
    at {\em the Workshop on Physics and
    Experiments with Linear $e^+ e^-$ Colliders}, Sep. 8--12, 1995,
    Morioka--Appi, Japan.
}} 
\author{
  Masaharu Tanabashi\footnote{E-mail: {\tt tanabash@theory.kek.jp}}, \quad
  Akiya Miyamoto\footnote{E-mail: {\tt miyamoto@kekux1.kek.jp}}}
\address{
  National Laboratory for High Energy Physics (KEK) \\
  Oho 1-1, Tsukuba 305, Japan}
\author{
  Ken-ichi Hikasa\footnote{E-mail: {\tt hikasa@tuhep.phys.tohoku.ac.jp}}}
\address{
  Department of Physics, Tohoku University\\
  Aoba-ku, Sendai 980-77, Japan}
\maketitle\abstracts{
  We discuss the sensitivity of the $e^+ e^- \rightarrow W^+
  W^-$  cross section at a future $e^+ e^-$ collider with
  $\sqrt{s}=500$GeV to the non-decoupling effects of 
  a techni-$\rho$ like vector resonance.
  The non-decoupling effects are parametrized by the chiral
  coefficients of the electroweak chiral perturbation theory. 
  We define renormalization scale independent chiral coefficients by
  subtracting the Standard Model loop contributions.
  We also estimate the size of the decoupling effects of the
  techni-$\rho$ resonance by using a phenomenological Lagrangian
  including the vector resonance.
}

\section{Introduction}
Chiral perturbation theory was originally introduced as
a systematic field theoretical method to parametrize low energy pion
physics and is given by a systematic expansion of chiral Lagrangian in
powers of derivatives and a consistent loop
expansion\cite{kn:We79b,kn:GL84}. 
It constructs the most general low energy 
pion scattering amplitude parametrized by chiral coefficients. 
The sizes of these chiral coefficients are known to be saturated by the
effects of heavier resonances\cite{kn:EGPR89}, i.e., $\rho$, $a_1$,
$\sigma$, etc.. 

If a new particle does not exist below the collider energy, then we
can use the same technique for the electroweak Higgs
sector\cite{kn:Ho91,kn:FLS91,kn:Wudka94}. 
The electroweak chiral Lagrangian parametrizes the most general form
of the non-decoupling effects in the Higgs sector. 
Electroweak chiral coefficients which are larger than the Standard
Model (SM) predictions might be a signal of the existence of TeV scale
new resonance states. 

So far, the sensitivity to these chiral coefficients at a future
$e^+e^-$ linear colliders has been discussed in their tree level
definition especially for the triple gauge boson
vertices\cite{kn:LHV95}.

In this talk we discuss the sensitivity to a techni-$\rho$ like
resonance at a future $e^+ e^-$ collider with $\sqrt{s}=500$GeV from 
the measurement of the electroweak chiral coefficients obtained from
the $W^+W^-$ cross section.
For such a purpose, we need to distinguish new physics from the SM
loop effects. 
We thus define renormalization scale independent chiral coefficients
by subtracting the SM loop contributions.
We calculate the sensitivity in a two dimensional plane of the triple
gauge boson vertex and the gauge boson two point functions, since
techni-$\rho$ contributes to both of them. 
We find the measurement of the chiral coefficients at the future $e^+
e^-$ collider with $\sqrt{s}=500$GeV is sensitive to a TeV scale
techni-$\rho$ like resonance, even though it cannot be observed 
directly. 

We also emphasize that, unlike the previous
studies\cite{kn:Barklow,kn:MHI94}, we do not use the equivalence
theorem.  
The effects of the one loop chiral logarithms can be taken into account
in our definition of the chiral coefficients. 
We can thus improve the previous calculation\cite{kn:CDDCDG} based on
the tree level BESS model\cite{kn:BESS}. 

We also discuss the size of decoupling effects for the case of the
relatively light techni-$\rho$ resonance.  

\section{The electroweak chiral Lagrangian}
We first review the chiral Lagrangian approach to electroweak symmetry
breaking.  
The chiral Lagrangian is constructed from the non-linearly realized
chiral field $U$
\begin{equation}
  U = \exp\left(\frac{i\tau^a w^a}{v}\right), \qquad 
  U \rightarrow g_L U g_Y^\dagger,
\end{equation}
where $g_L$ and $g_Y$ are 
\begin{equation}
  g_L = \exp \left(i\frac{\tau^a}{2}\theta_L^a\right) \in SU(2)_L,
  \quad 
  g_Y = \exp \left(i\frac{\tau^3}{2}\theta_Y \right) \in U(1)_Y,
\end{equation}
with $w^a$ are the would-be Nambu-Goldstone fields and 
$v\simeq 250$GeV is the vacuum expectation value.

The chiral Lagrangian can be expanded in terms of the chiral
dimension, i.e., the number of derivatives.  We consider here
operators through $O(\partial^4)$, since coefficients of higher
dimensional operators are suppressed by the mass scale of new
particles. 

The electroweak chiral Lagrangian at $O(\partial^2)$ is given by
\begin{equation}
  {\cal L}_2 = -\frac{v^2}{4}\mbox{tr}(V_\mu V^\mu) 
         +\beta_1 \frac{v^2}{4} \mbox{tr}(V_\mu T)\mbox{tr}(V^\mu T),
\label{eq:chiral2}
\end{equation}
where $V_\mu$ and $T$ are given by
\begin{equation}
  V_\mu   \equiv D_\mu U \cdot U^\dagger,
  \quad
  T       \equiv U \tau^3 U^\dagger,
  \qquad
  D_\mu U \equiv \partial_\mu U 
            + i g_2 W_\mu^a \frac{\tau^a}{2} U 
            - i g_Y B_\mu U \frac{\tau^3}{2}. 
\end{equation}
The chiral coefficient $\beta_1$ corresponds to the $\rho$ parameter
$\rho-1=\Delta\rho \simeq 2\beta_1$.  

Assuming $CP$ invariance in the Higgs sector, we find 
eleven independent operators at the $O(\partial^4)$ level. 
We follow the notation of Appelquist and Wu\cite{kn:Lo80,kn:AW93}:
\begin{eqnarray}
  {\cal L}_4
  &=& \alpha_1 g_2 g_Y \tr({\cal W}^{\mu\nu} U {\cal B}_{\mu\nu} U^\dagger) 
    +i\alpha_2 g_Y \tr(U^\dagger [V_\mu,V_\nu] U {\cal B}^{\mu\nu}) 
    +i\alpha_3 g_2 \tr([V_\mu,V_\nu] {\cal W}^{\mu\nu}) 
  \nonumber\\
  & &+\alpha_4 \tr(V_\mu V_\nu)\tr(V^\mu V^\nu)
     +\alpha_5 \tr(V_\mu V^\mu)\tr(V_\nu V^\nu)
  \nonumber\\
  & &+\alpha_6 \tr(V_\mu V_\nu) \tr(T V^\mu)\tr(T V^\nu)
     +\alpha_7 \tr(V_\mu V^\mu) \tr(T V_\nu)\tr(T V^\nu)
  \nonumber\\
  & &+\frac{1}{4}\alpha_8 g_2^2 \tr(T {\cal W}_{\mu\nu})
                                \tr(T {\cal W}^{\mu\nu})
     +\frac{i}{2} \alpha_9 g_2 \tr(T {\cal W}_{\mu\nu})
                               \tr(T [V^\mu,V^\nu])
  \nonumber\\
  & &+\frac{1}{2} \alpha_{10} \tr(T V_\mu) \tr(T V^\mu) 
                              \tr(T V_\nu) \tr(T V^\nu)  
     +\alpha_{11} g_2 \epsilon^{\mu\nu\rho\lambda}
     \tr(T V_\mu)\tr(V_\nu {\cal W}_{\rho\lambda}),
\label{eq:AW}
\end{eqnarray}
with $\epsilon_{0123}=-\epsilon^{0123}=1$.
The operators $\alpha_1$ and $\alpha_8$ lead to non-minimal two points
gauge boson vertices and correspond to $S\simeq -16\pi\alpha_1$ and $U\simeq 
-16\pi\alpha_8$ parameters\cite{kn:PT90}, respectively.   
The operators $\alpha_{i=1,2,3,8,9,11}$ correspond to anomalous triple
gauge vertices which we will investigate in this talk.
$\alpha_{i=3,4,5,6,7,8,9,11}$ correspond to non-minimal quadruple
gauge vertices\cite{kn:Mi95}.
We also note that the operators $\alpha_{i=6,7,8,9,10,11}$ violate the
custodial symmetry, and thus the sizes of these coefficients are
expected to be smaller than the others. 

One loop diagrams of the $O(\partial^2)$ Lagrangian of
Eq.(\ref{eq:chiral2}) also contribute to the $O(\partial^4)$ amplitudes. 
The logarithmic divergences of these diagrams are absorbed by 
redefinitions of the $O(\partial^4)$ chiral coefficients $\alpha_i$, 
\begin{equation}
  \alpha_i \rightarrow \alpha_i^r(\mu),
\end{equation}
with $\alpha_i^r(\mu)$ being renormalized at the scale $\mu$.  
We follow the renormalization scheme of Gasser and
Leutwyler\cite{kn:GL84}. 

We define renormalization scale independent chiral coefficients by
subtracting the SM contribution,
\begin{equation}
  \hat \alpha_i \equiv \alpha_i^r(\mu) - \alpha_{i,{\rm SM}}^r(\mu).
\end{equation}
Calculating the matching condition of the chiral perturbation theory
and the one doublet Higgs model at $p^2\ll M_H^2$, we obtain the SM
contributions to the chiral coefficients\cite{kn:GL84};
\begin{eqnarray}
\alpha_{1,{\rm SM}}^r(\mu) 
  &=& -\frac{1}{12(4\pi)^2} \left[ 
          \frac{11}{6} -\ln\left(\frac{\mu^2}{M_H^2}\right)
       \right],
\label{eq:sm1}
  \\
\alpha_{2,{\rm SM}}^r(\mu)
  &=&  \alpha_{3,{\rm SM}}^r(\mu)
   =   \frac{1}{24(4\pi)^2} \left[ 
          \frac{11}{6} +\ln\left(\frac{\mu^2}{M_H^2}\right)
       \right].
\label{eq:sm23}
\end{eqnarray}
We note here that the Higgs mass, $M_H$, is introduced as an artificial
parameter for the definition of the renormalization scale invariant
chiral coefficients.

\section{Form factors}
The $e^+ e^-\rightarrow W^+ W^-$ process is sensitive both to the
gauge boson two point functions and to the triple gauge boson vertices.
We first consider the $e^+ e^- \rightarrow f\bar f$ process to clarify 
the structure of the gauge boson two point functions.

The amplitude of the $e^+ e^- \rightarrow f\bar f$ process with
oblique correction is given by\cite{kn:KL89}
\begin{equation}
  {\cal M}(e^+ e^- \rightarrow f\bar f)
     = e_*^2 \frac{QQ'}{-p^2} 
            + \frac{e_*^2}{s_*^2 c_*^2}
              \frac{(I_3-s_*^2 Q)(I_3'-s_*^2 Q')}{-p^2 + M_{Z*}^2},
\end{equation}
where $e^2_*(p^2)$, $s^2_*(p^2)$ and $M^2_{Z*}(p^2)$ are functions of 
momentum. 

The power type running of dimensionless functions
$e_*$ and $s_*$ is suppressed by the mass scale of the new particles.
On the other hand, $M^2_{Z*}$ can have power type running
(a non-decoupling effect).
We also note that the logarithmic running of these functions ($e_*$,
$s_*$, $M_{Z*}$) is determined solely from their imaginary parts via  
dispersion relations. 
We can thus determine the whole structure of these functions below the
threshold of new particles:
\begin{eqnarray}
  M_{Z*}^2(p^2) 
    &=& M_{Z*}^2(p^2)_{\rm SM} 
       - \frac{e^2}{s^2 c^2} (p^2-M_Z^2) \hat\alpha_1, 
\label{eq:mzs}  \\
  s_*^2(p^2)
    &=& s_*^2(p^2)_{\rm SM},
\label{eq:ss}
 \\
  e_*^2(p^2)
    &=& e_*^2(p^2)_{\rm SM},
\label{eq:es}
\end{eqnarray}
where the SM form factors $M_{Z*}^2(p^2)_{\rm SM}$,
$s_*^2(p^2)_{\rm SM}$ and $e_*^2(p^2)_{\rm SM}$ are 
calculated using  ($M_Z$,$e_*(M_Z^2)$,$s_*(M_Z^2)$) as a set of input 
parameters.\footnote{
We take this less familiar renormalization scheme to simplify
our calculation.  It is also possible to take the standard 
renormalization scheme using ($G_\mu$, $\alpha_{QED}$, $M_Z$) as a set 
of input parameters.  
In this case, however, the analogues of 
Eqs.(\ref{eq:mzs})--(\ref{eq:es}) and
Eqs.(\ref{eq:f3s})--(\ref{eq:f5s}) become more 
complicated\cite{kn:HIMT96}.
}
It should be noted that $\alpha_1$ can be measured from neutral
current quantities, while we need information of charged current
(e.g., the muon decay constant, $G_\mu$,  
and the $W$ boson mass) for the determination of the other oblique
parameters ($\beta_1\simeq\Delta\rho/2$, $\alpha_8\simeq-U/(16\pi)$).
\begin{figure}[htbp]
  \begin{minipage}[t]{0.38\textwidth}
  \begin{center}
    \leavevmode
    \psfig{file=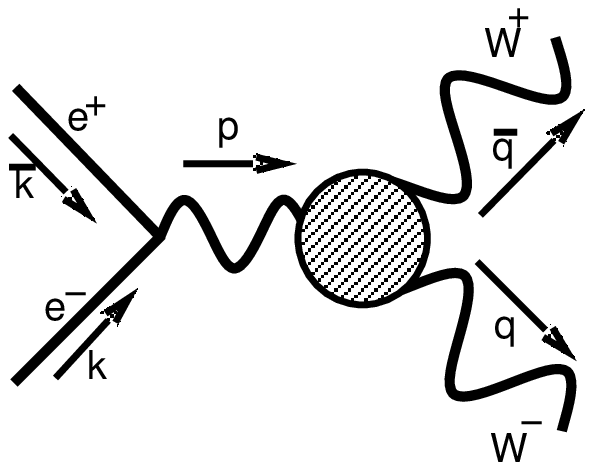,width=2in}
    \caption{
       The $s$-channel amplitude of the $e^+ e^-\rightarrow W^+ W^-$ process.}
    \label{fig:eeww}
  \end{center}
  \end{minipage}
  \hfil
  \begin{minipage}[t]{0.57\textwidth}
  \begin{center}
    \leavevmode
    \psfig{file=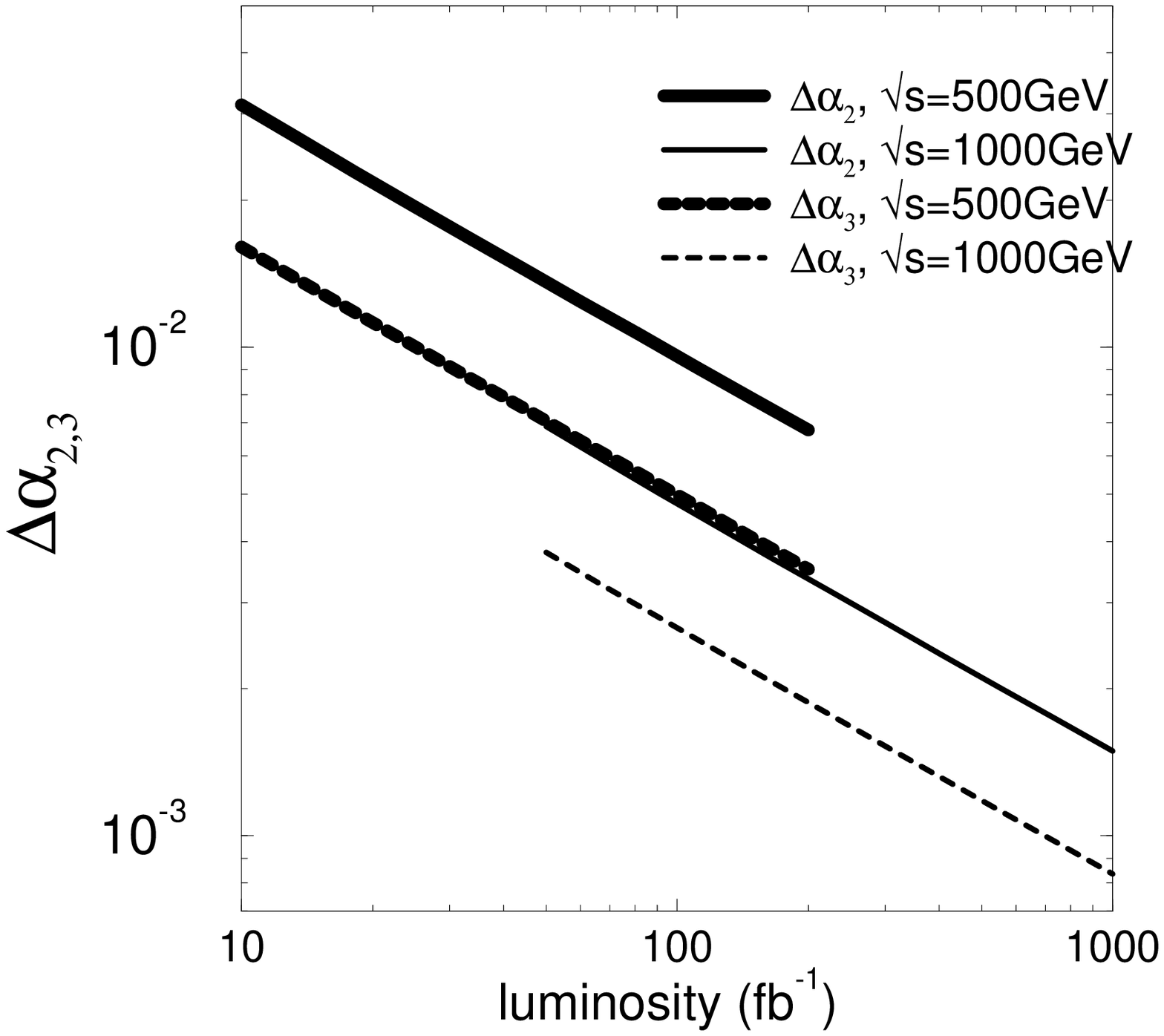,width=3in}
    \caption{
       The statistical errors of $\alpha_2$ and $\alpha_3$ for
       $\Delta\chi^2=4$ at the future 
       $e^+e^-$ collider with $\protect\sqrt{s}=$500GeV and 1000GeV.
}
    \label{fig:a23}
  \end{center}
  \end{minipage}
\end{figure}

We are now ready to discuss the $e^+ e^- \rightarrow W^+ W^-$ process. 
The corresponding amplitude can be written as
\begin{equation}
  {\cal M}(e^+ e^- \rightarrow W^+ W^-) = {\cal M}_S + {\cal M}_T, 
\end{equation}
with ${\cal M}_S$ and ${\cal  M}_T$ being $s$-channel and $t$-channel
amplitudes respectively. The $s$-channel amplitude can be written as
(see Fig.\ref{fig:eeww})
\begin{equation}
{\cal M}_S = 
  e_*^2 \gamma^\mu Q \frac{1}{-p^2}\Gamma_{\mu\alpha\beta}^{\gamma *}
 +\frac{e_*^2}{s_*^2} \gamma^\mu (I_3-s_*^2 Q) 
  \frac{1}{-p^2 + M_{Z*}^2} \Gamma_{\mu\alpha\beta}^{Z *}.
\end{equation}
The $WWV$ vertices, $\Gamma_{\mu\alpha\beta}^{\gamma*}$ and
$\Gamma_{\mu\alpha\beta}^{Z*}$, can be expressed in terms of the form
factors\cite{kn:HPZH}:   
\begin{eqnarray}
  \Gamma_{\mu\alpha\beta}^{V*}
  &=& f_1^{V*}(q-\bar q)_\mu g_{\alpha\beta}
     -f_2^{V*}(q-\bar q)_\mu p_\alpha p_\beta
     +f_3^{V*}(p_\alpha g_{\mu \beta}-p_\beta g_{\mu\alpha})
  \nonumber\\
  & &+if_5^{V*}\epsilon_{\mu\alpha\beta\rho}(q-\bar q)^\rho.  
\end{eqnarray}
The form factors $f_3^\gamma$, $f_1^Z$, $f_3^Z$ and $f_5^Z$ 
depend on the chiral coefficients $\alpha_{1,2,3,8,9,11}$,
\begin{eqnarray}
f_3^{\gamma *} 
  &=& f_3^{\gamma *}{}_{\rm SM} 
     + \frac{e^2}{s^2} (-\hat\alpha_1 +\hat\alpha_2 +\hat\alpha_3
     -\hat\alpha_8+\hat\alpha_9 ),
\label{eq:f3s}
\\
f_1^{Z *} 
  &=& f_1^{Z *}{}_{\rm SM}
     + \frac{e^2}{s^2c^2}\hat\alpha_3,
\\
f_3^{Z *} 
  &=& f_3^{Z *}{}_{\rm SM} 
  + \frac{e^2}{s^2c^2}\hat\alpha_3
  + \frac{e^2}{c^2} (\hat\alpha_1 - \hat\alpha_2)
          + \frac{e^2}{s^2} (\hat\alpha_3 - \hat\alpha_8 + \hat\alpha_9), 
\\
f_5^{Z *}
  &=& f_5^{Z *}{}_{\rm SM} + \frac{ e^2 }{s^2 c^2} \hat\alpha_{11},
\label{eq:f5s}
\end{eqnarray}
while $f_1^\gamma$, $f_2^\gamma$, $f_5^\gamma$ and $f_2^Z$ do not
receive non-decoupling effects from heavy particles.
The $\ln M_H$ dependence cancels between $f^{V*}_{i,{\rm SM}}$ and
$\hat\alpha_i$. 
The remaining  $M_H^2$ dependence in $f^*_{SM}$ is suppressed by $1/M_H^2$.

In a similar manner, the $t$-channel neutrino exchange amplitude is
given by: 
\begin{equation}
{\cal M}_T 
  = \frac{e_*^2(t)}{s_{W*}^2(t)} 
    \gamma_\alpha 
         \frac{1-\gamma_5}{2} \frac{1}{\fsl{q}-\fsl{k}} 
    \gamma_\beta, \qquad t \equiv (q-k)^2,
\end{equation}
with $s_{W*}(t)$ being
\begin{equation}
  s^2_{W*}(t) = s^2_{W*}(t)_{\rm SM} + e^2 (\hat\alpha_1 + \hat\alpha_8).
\end{equation}

We are now ready to evaluate the sensitivity limit to these
chiral coefficients of the $e^+ e^- \rightarrow W^+ W^-$
differential cross section.
The angular distribution can be measured from the decay
$WW\rightarrow \ell\nu,q\bar q$, which has a 28\% branching fraction.
We use the differential cross section in the range $-0.8<\cos\theta<0.8$. 
A detection efficiency of 50\% is assumed for the decay $WW\rightarrow 
\ell\nu,q\bar q$.

In the sensitivity limit calculation,
we can neglect the SM loop contribution in the running of the form
factors.
We also neglect the uncertainty of the SM input parameters.
As a set of input parameters of the SM we use 
($M_Z=91.19$GeV, $4\pi/e^2_*(M_Z^2)=128.72$, $s^2_*(M_Z^2)=0.2305$). 

Fig.\ref{fig:a23} shows the sensitivity
($\Delta\chi^2=\chi^2-\chi^2_{\rm min}<4$) to the chiral coefficients 
$\alpha_{2,3}$ as functions of the integrated luminosity of a future 
$e^+ e^-$ collider.
When making the graph of $\Delta\alpha_2$ ($\Delta\alpha_3$), we
assumed that all chiral coefficients other than $\alpha_2$
($\alpha_3$) are zero. 
We discuss physical meaning of this sensitivity in the next section.


\section{A model of techni-$\rho$ like resonance}
We next evaluate the size of the chiral coefficients
$\alpha_{1,2,3,8,9,11}$ predicted in a techni-$\rho$ like vector
resonance model.  
For such a purpose we first construct a phenomenological Lagrangian of 
the vector resonance.

One of the most familiar chiral Lagrangian formulations of the vector
resonance is the hidden local symmetry formalism\cite{kn:BKUYY}.
The usual phenomenological Lagrangian with hidden local symmetry
contains two independent parameters, $a$ and $g$.
The techni-$\rho$ like resonance has three independent observable
quantities when it is on-shell (total decay width, fermionic decay
width, and its mass). 
We thus need to generalize the hidden local symmetry Lagrangian:
\begin{equation}
  {\cal L} 
  = v^2  \mbox{tr}(\hat\alpha_{\mu\perp}\hat\alpha^\mu_\perp)
     +a v^2 \mbox{tr}(\hat\alpha_{\mu\parallel}\hat\alpha^\mu_\parallel)
     -\frac{1}{2g^2} \mbox{tr}(V_{\mu\nu}V^{\mu\nu})
    +\frac{z_4}{2} i \mbox{tr}
      (V_{\mu\nu} [\hat\alpha^\mu_\perp, \hat\alpha^\nu_\perp]),
\label{eq:hls}
\end{equation}
where $V_\mu$ stands for the techni-$\rho$ resonance field.  
The Maurer-Cartan one forms $\hat\alpha_{\mu\perp}$ and
$\hat\alpha_{\mu\parallel}$ are defined by
\begin{equation}
  \hat\alpha_{\mu\perp} 
  = \frac{1}{2i}\left[ 
         D_\mu \xi_L\cdot \xi_L^\dagger
        -D_\mu \xi_R\cdot \xi_R^\dagger
      \right], \qquad
  \hat\alpha_{\mu\parallel} 
  = \frac{1}{2i}\left[ 
         D_\mu \xi_L\cdot \xi_L^\dagger
        +D_\mu \xi_R\cdot \xi_R^\dagger
      \right],
\end{equation}
with $\xi_L$, $\xi_R$ from $U=\xi_L^\dagger \xi_R$.
The covariant derivative $D_\mu$ is given by
\begin{equation}
  D_\mu \xi_L 
  = \partial_\mu \xi_L - iV_\mu \xi_L 
      - ig_2 \xi_L W_\mu^a \frac{\tau^a}{2}, \qquad
  D_\mu \xi_R 
  = \partial_\mu \xi_R - iV_\mu \xi_R 
      - ig_Y \xi_R B_\mu \frac{\tau^3}{2}.
\end{equation}

In addition to the usual parameters of the hidden local symmetry
Lagrangian, $a$ and $g$, we introduced one additional parameter,
$z_4$, which parametrizes the non-minimal coupling of the vector
resonance. 
We can show the equivalence of this formulation with the anti-symmetric 
tensor formulation.\cite{kn:Tana95}
The relation to the BESS model will be clarified
elsewhere\cite{kn:HIMT96}. 

The mass of the techni-$\rho$ resonance is given by
\begin{equation}
  M_V^2 = g^2 a v^2. 
\end{equation}
In the QCD-like technicolor model, vector meson dominance and the
KSRF relation\cite{kn:KS66} lead to the parameters 
\begin{equation}
  a=2, \qquad z_4=0. 
\end{equation}

We next consider the matching of the electroweak chiral Lagrangian 
of Eq.(\ref{eq:AW}) with the phenomenological vector resonance model
Eq.(\ref{eq:hls}).
We assume that the tree level matching conditions can be applied at
the scale of the techni-$\rho$ resonance,
\begin{eqnarray}
  \alpha_1^r(\mu=M_V) 
  &=& -\frac{a v^2}{4M_V^2}, 
\label{eq:treea1}
\\ 
  \alpha_2^r(\mu=M_V) 
  &=& \alpha_3^r(\mu=M_V) 
   =  -\frac{a v^2}{8M_V^2} + \frac{z_4}{16}.
\label{eq:treea23}
\end{eqnarray}
This assumption is known to work well in the case of the low energy
pion chiral Lagrangian\cite{kn:EGPR89}. 

Subtracting the SM contribution of Eqs.(\ref{eq:sm1})--(\ref{eq:sm23})
from Eqs.(\ref{eq:treea1})--(\ref{eq:treea23}), we find
\begin{eqnarray}
  \hat\alpha_1 &=& -\frac{a v^2}{4M_V^2} 
               +\frac{1}{12(4\pi)^2}\left[
                  \frac{11}{6} - \ln\left(\frac{M_V^2}{M_H^2}\right)
                \right], 
\label{eq:a1}
  \\
  \hat\alpha_2 &=& \hat\alpha_3 
            = -\frac{a v^2}{8M_V^2} + \frac{z_4}{16}
               -\frac{1}{24(4\pi)^2}\left[
                  \frac{11}{6} + \ln\left(\frac{M_V^2}{M_H^2}\right)
                \right],
\label{eq:a23}
  \\
  \hat\alpha_8 &=& \hat\alpha_9 = \hat\alpha_{11} = 0.
\end{eqnarray}
The logarithmic 
correction $\ln(M_V)$ is due to the renormalization group evolution of
the chiral coefficients below the mass of the techni-$\rho$ resonance.

\begin{figure}[htbp]
  \begin{minipage}{\textwidth}
  \begin{minipage}[t]{0.48\textwidth}
  \begin{center}
    \leavevmode
    \psfig{file=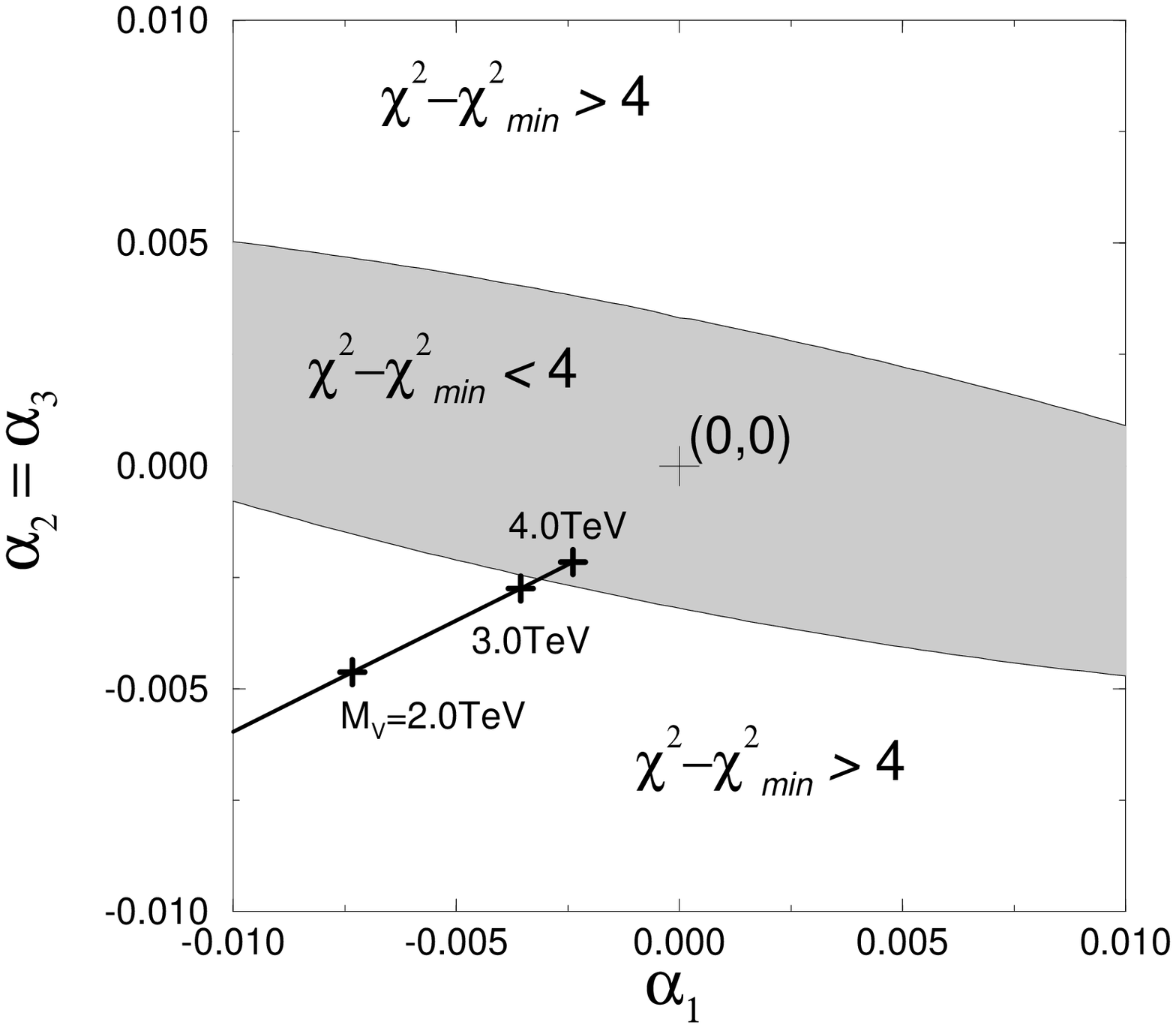,width=2.7in}
    \caption{
       The statistical errors in the $\alpha_1$--$(\alpha_2=\alpha_3)$ plane
       for $\Delta\chi^2=4$ at a future $e^+e^-$ 
       collider with $\protect\sqrt{s}=$500GeV and $\int{\cal
         L}=100$fb$^{-1}$. 
       The techni-$\rho$ contribution is also depicted for the QCD-like
       technicolor model.} 
    \label{fig:a1a23}
  \end{center}
  \end{minipage}
  \hfil
  \begin{minipage}[t]{0.48\textwidth}
  \begin{center}
    \leavevmode
    \psfig{file=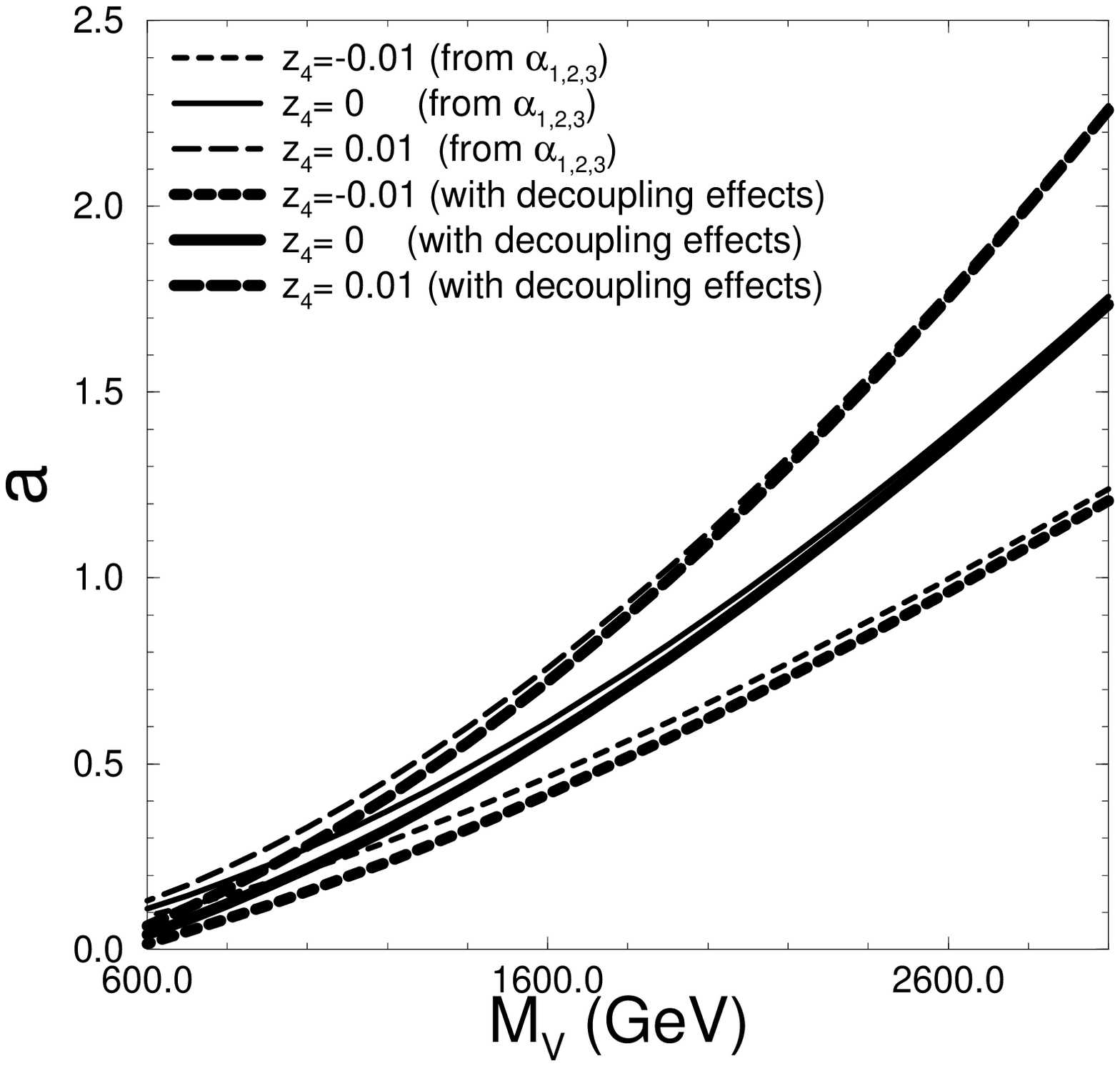,width=2.5in}
    \caption{
      The vector resonance mass lower bound obtained from the
      electroweak chiral coefficients. 
      The region above the curves is excluded.
      Constraints including the decoupling effects are also shown.}
    \label{fig:tcrho}
  \end{center}
  \end{minipage}
  \end{minipage}
\end{figure}

Since the techni-$\rho$ resonance contributes both to $\alpha_1$ and
to $\alpha_2=\alpha_3$, we need to calculate the limit contour in the
$\alpha_1$--($\alpha_2=\alpha_3$) plane.
The limit contour for $\Delta\chi^2 = \chi^2 - \chi^2_{\rm
  min}<4$ is shown in Fig.\ref{fig:a1a23} for $\sqrt{s}=500$GeV and
an integrated luminosity of $\int{\cal L}=100$fb$^{-1}$ assuming
$\chi^2_{\rm min}$ corresponds to the SM with $M_H=1$TeV.
In Fig.\ref{fig:a1a23} the techni-$\rho$ contribution 
Eqs.(\ref{eq:a1})--(\ref{eq:a23}) for a QCD-like
technicolor model $a=2,z_4=0$ is also depicted. 
The limit of Fig.\ref{fig:a1a23} corresponds to $M_V\gtrsim 3$TeV for
the QCD-like techni-$\rho$ resonance.
We summarize in Fig.\ref{fig:tcrho} the sensitivity to the 
techni-$\rho$ resonance for the generalized parameter space of the
vector resonance model. 

So far we have considered non-decoupling effects and neglected the
decoupling corrections.
We need to be careful for the case of a light vector resonance, however, 
since decoupling effects may play an important role.
For such a purpose we calculate the form factors in the
techni-$\rho$ resonance model without making the momentum expansion.
Fig.\ref{fig:tcrho} shows the sensitivity limit calculated from these
form factors including the decoupling corrections.   
We find that the decoupling effects are negligible over a wide range of
parameters. 

We note that the uncertainties of the SM input parameters and the
luminosity measurements are neglected in this talk.
We should also combine our analysis with LEP/SLC precision
measurements for a detailed study.
The analysis with respect to  these problems will be published
elsewhere\cite{kn:HIMT96}. 

\section{Summary}

We have determined the sensitivity of the $e^+ e^-\rightarrow W^+ W^-$
at a future $e^+ e^-$ linear collider to non-decoupling effects
(electroweak chiral coefficients).
The renormalization scale independent electroweak chiral coefficients
are defined by subtracting the SM contributions.
The effect of one loop chiral logarithms can be taken into account in
this definition of the chiral coefficients.
A future $e^+ e^-$ collider with $\sqrt{s}=500$GeV, $\int{\cal
  L}=100$fb$^{-1}$ can measure these chiral coefficients up to the
statistical errors $\Delta\alpha_2\simeq 0.01$ and
$\Delta\alpha_3\simeq 0.005$  for $\Delta\chi^2=4$. 

The sensitivity to the techni-$\rho$ like resonance can be extracted 
from this analysis.
The estimated statistical error in the 
$\hat\alpha_1$--($\hat\alpha_2=\hat\alpha_3$) plane corresponds to a
sensitivity to a techni-$\rho$ with a mass $M_V\simeq 3$TeV for the
QCD-like technicolor model assuming that the $\chi^2_{\rm min}$
corresponds to the SM with $M_H=1$TeV. 
The decoupling effects of the vector resonance are also investigated
and found to be negligible over a wide range of parameters.

\section*{Acknowledgements}
The authors thank Y. Okada, M.M. Nojiri and R. Szalapski for careful
reading of the manuscript.

\end{document}